\documentclass[11pt]{article}
\usepackage[utf8]{inputenc}
\usepackage{amsmath}
\usepackage{amssymb}
\usepackage{color}
\usepackage{graphicx}
\usepackage{hyperref}
\usepackage{float}
\usepackage{cite}
\usepackage{esint}
\usepackage{babel}
\usepackage{mathdots}
\usepackage{amsfonts}
\usepackage{mathrsfs}

\def \be {\begin{equation}}
\def \ee {\end{equation}}
\def \nn {\nonumber}
\usepackage[dvipsnames]{xcolor}
\newcommand{\VF}[1]{}
\newcommand{\JW}[1]{}

\begin{document}

\author{Hong-Zhe Zhang\footnote{Email: 2020210009@tju.edu.cn.},~~Wan-Zhe Feng\footnote{Email: vicf@tju.edu.cn, corresponding author.},~~Jun-Bao Wu\footnote{Email: junbao.wu@tju.edu.cn, corresponding author.}\\
\textit{\small Center for Joint Quantum Studies and Department of Physics,}\\
\textit{\small School of Science, Tianjin University, Tianjin 300350, PR. China}}

\title{\LARGE Holographic Operator Product Expansion of Loop Operators in $\mathcal{N}=4$ $SO(N)$ Super Yang-Mills Theory}

\date{}
\maketitle

\begin{abstract}

In this paper, we compute the correlation functions of Wilson(-'t~Hooft) loops with chiral
primary operators in
$\mathcal{N}=4$ supersymmetric Yang-Mills theory with $SO(N)$ gauge symmetry, which has a holographic dual description of Type IIB superstring theory on the
$AdS_{5}\times\mathbf{RP}^{5}$ background.
Specifically, we compute the coefficients of the chiral primary operators in the operator product expansion of
Wilson loops in the fundamental representation, Wilson-'t Hooft loops in the symmetric representation, Wilson loops in the anti-fundamental representation and the spinor representation. We also compare these results to the $\mathcal{N}=4$ $SU(N)$ super Yang-Mills theory.

\end{abstract}

\newpage

\tableofcontents

\section{Introduction}\label{introduction}

The holographic duality between the maximally supersymmetric Yang-Mills theory (SYM) with $SU(N)$ gauge
group and Type IIB string theory on the $AdS_{5}\times S^{5}$ background
is a most studied  example of the AdS/CFT correspondence~\cite{Maldacena:1997re}.
 The vacuum expectation values
of Wilson loops are natural observables in gauge theories and they
are also calculable from the AdS side. In the string theory
description, a Wilson loop\footnote{More precisely speaking, here we mean the Wilson-Maldacena loop~\cite{Maldacena:1998im, Rey:1998ik} whose definition involves the scalars in the $\mathcal{N}=4$ SYM as well.} in the fundamental representation are related to a fundamental string  whose worldsheet ending
on the AdS boundary along the contour of this Wilson loop~\cite{Maldacena:1998im, Rey:1998ik}. The on-shell action, with the boundary terms from Legendre transformation included~\cite{Drukker:1999zq}, give the prediction for the vacuum expectation value ({\it vev}) of this Wilson loop at large $N$ and large 't~Hooft coupling $\lambda\equiv g_{{\rm YM}}^{2}N$,
in which case the classical string theory is a good approximation
with large string tension and small curvature. This holographic prediction matches with the  field theory results in the large $N$ and $\lambda$ limit. The field theory results were obtained  based on the conjecture that the computations  can  be reduced to the ones in a  Gaussian matrix model~\cite{Erickson:2000af}. Later this conjecture was proved using supersymmetric localization~\cite{{Pestun:2007rz}}.  This match provided a highly non-trivial check to the AdS/CFT conjecture since the {\it vev} of the Wilson loop
is a non-trival function of $\lambda$ and $N$.  Higher rank Wilson
loops in gauge theories are
dual to  D-branes carrying electric
flux on the their worldvolume~\cite{Drukker:2005kx, Yamaguchi:2006tq, Gomis:2006im, Gomis:2006sb}.
When the rank of the representation is sufficiently large,
the back reaction from the D-branes must be considered.
The Wilson loop in the higher rank representation with mixed symmetries
is dual to a certain bubbling supergravity solution~\cite{Yamaguchi:2006te,  Lunin:2006xr, DHoker:2007mci}.  We will not discuss such supergravity solutions in this paper.

Specifically, half-BPS circular Wilson loops in the rank-$k$ symmetric representation
of the gauge group corresponds to D3-brane with $AdS_{2}\times S^{2}$
worldvolume and $k$ units of fundamental string charge~\cite{Drukker:2005kx}. While half-BPS circular
Wilson loops in the rank-$k$ anti-symmetric representation of the
gauge group have a bulk description in terms of $AdS_{2}\times S^{4}$
D5-brane with $k$ units of fundamental string charge~\cite{Yamaguchi:2006tq}. These D-branes
are $1/2$-BPS and preserve the same $SO(2,1)\times SO(3)\times SO(5)$
isometries. On the other hand, the 't Hooft loop, which is the magnetic
dual of Wilson loop, can be obtained using S-duality in $\mathcal{N}=4$ SYM. General  $SL(2, \mathbf{Z})$  transformation maps a Wilson loop to a Wilson-'t Hooft (WH) loop~\cite{Kapustin:2005py}.
It was proposed in~\cite{Chen:2006iu} that a WH loop in symmetric representations of both  the gauge group and its Goddard-Nuyts-Olive (GNO) dual group~\cite{Goddard:1976qe}  (the Langlands dual group) is dual to a  D3-brane carrying both F-string and D-string charges. More details of such WH loops will be provided later in this section.

 A circular Wilson loop
can be expanded in a series of local operators with
different conformal dimensions, when the probing distance is much larger than the radius of this loop.
Half-BPS chiral primary operators (CPOs) are an important
class of operators with protected dimensions appearing in this operator product expansion (OPE). The OPE coefficient   can   be extracted from the correlation function of the Wilson loop and the local operators~\cite{Berenstein:1998ij}. In the large $N$ and $\lambda$ limit, the correlation
function of a Wilson loop in the fundamental representation with a CPO can be derived by calculating the coupling of the supergravity modes dual to this CPO to the string worldsheet~\cite{Berenstein:1998ij}.
Similar procedure can be used to compute   the correlator of a higher rank Wilson loop with  a CPO
using D3$_{k}$ and D5$_{k}$
branes and replacing string worldsheet by the brane worldvolume~\cite{Giombi:2006de}. These results were confirmed by the results in the field theory side using the  matrix model~\cite{Semenoff:2001xp,Giombi:2006de}.
The reduction to this matrix model computations was later proved by supersymmetric localization~\cite{Giombi:2009ds}.

The $\mathcal{N}=4$ SYM theory with gauge group $SO(N)$ has some features different from the $SU(N)$ theory. When $N$ is odd, the group is non-simply-laced and the S-dual theory
has gauge algebra $sp(\frac{N-1}2)$~\cite{Goddard:1976qe}.
In this case, the gauge algebras before and after the S-duality transformation are different. This is distinct from the S-duality transformation of  the theory with gauge group $SU(N)$. When $N$ is even, the group $SO(N)$ is simply-laced and the dual theory still has the gauge  algebra $spin(N)$.
Another notable feature regarding the Wilson loops in $SO(N)$ theories is the presence of Wilson loops in spinor representations.

In string theory, ${\mathcal{N}}=4$~$SO(N)$ SYM can be realized as the low energy effective theory of coincident D3-branes on top of  a suitable O3 plane. Based on this,  Witten proposed that  the ${\mathcal{N}}=4$~$SO(N)$ SYM  is holographically  dual to string theory on the $AdS_{5}\times\mathbf{RP}^{5}$
orientifold~\cite{Witten:1998xy}. The five-dimension real projective space $\mathbf{RP}^{5}$
is obtained by the five-dimensional sphere $S^{5}$ by identifying antipodal points, $\mathbf{RP}^5=S^5/\mathbf {Z}_2$.
This correspondence was recently studied in~\cite{Giombi:2020kvo}.
It has been demonstrated that the expectation value of the Wilson loop in the spinor representation of the gauge group, calculated through supersymmetric localization~\cite{Fiol:2014fla, Giombi:2020kvo}, precisely matches the result obtained from the D5-brane, with its worldvolume including the $\mathbf{RP}^{4}$ subspace of $\mathbf{RP}^{5}$.
The holographic description of Wilson loops in the fundamental, symmetric, and anti-symmetric representations were also studied, and the holographic prediction of their {\it{vev}}s matches exactly with the result from supersymmetric localization~\cite{Fiol:2014fla, Giombi:2020kvo}. In this
paper, we  compute the correlation functions of Wilson(-'t~Hooft) loops with chiral
primary operators of $\mathcal{N}=4$ SYM with $SO(N)$ gauge symmetry.
The line operators we will consider include:
\begin{itemize}
  \item The half-BPS circular Wilson loop in the fundamental representation of Lie algebra $g=spin(N)$, $W_\Box$.
   \item The half-BPS circular Wilson loops in the $k$-th anti-symmetric representation of $g$, $W_{A_k}$.
   \item The half-BPS circular Wilson loops in the spinor representation of $g$, $W_{sp}$.
  \item Special half-BPS circular WH loops. Recall that  WH loops \cite{Kapustin:2005py} are labelled by\\
    $(\lambda_{elec.}, \lambda_{mag.})\in \Lambda_w\times \Lambda_{mw}$ with the identification \be (\lambda_{elec.}, \lambda_{mag.})
   \sim (w\lambda_{elec.}, w\lambda_{mag.}), w\in W.\ee Here $\Lambda_w$ and $\Lambda_{mw}$
  are weight lattices of $g$ and $^Lg$, respectively. $^L g$ is the GNO dual group~\cite{Goddard:1976qe} of $g$  \footnote{When $g=spin(2n^\prime)$, $^Lg=spin(2n^\prime)$. And when $g=spin(2n^\prime+1)$, $^Lg=sp(n^\prime)$. Here $n^\prime=N/2$ for even $N$ and $n^\prime=(N-1)/2$ for odd $N$.},
$W$ is the Weyl group of $g$ and $^Lg$. We focus on the case that the $W$-orbit $[\lambda_{elec.}]$ corresponds to the $n$-th symmetric representation of $g$ and the $W$-orbit $[\lambda_{mag.}]$ corresponds to the  $m$-th symmetric  representation of $^Lg$.  We label these WH loops by ${WH}_{S_n, S_m}$'s.
      \end{itemize}

The paper is organized as follows. In Section~\ref{Sec:2}~and~\ref{Sec:3},
we will briefly review the dual string description of the ${\mathcal{N}}=4$~$SO(N)$ theory and the half-BPS CPOs with their gravity duals.  In Section~\ref{fudamental}, \ref{symmetric}, \ref{antisymmetric} and \ref{spinor}, we will compute the OPE coefficients of these CPOs
in the OPE expansion of the Wilson loops in the fundamental representation, the WH loops in the symmetric representation, the Wilson loops in the anti-fundamental representation and the spinor representation, respectively. The final section is devoted to the conclusion and the discussions.
In  Appendix~\ref{appendix}, we briefly discuss the coefficient of the bulk-to-boundary propagator of a certain mode in $AdS_5$.

\section{The string theory description of the $\mathcal{N}=4$ $SO(N)$ theory}\label{Sec:2}
Four-dimensional $\mathcal{N}=4$ SYM with gauge group $SO(N)$ is dual to Type IIB superstring theory on the $AdS_5\times \mathbf{RP}^5$ background with Ramond-Ramond (RR) $5$-form fluxes $F_5$~\cite{Witten:1998xy}.  We should make also a choice of ``discrete torsion'' of RR $2$-form $B_{RR}$. We will describe this discrete torsion later. In the large $N$ and large 't~Hooft coupling limit, the IIB supergravity on $AdS_5\times \mathbf{RP}^5$ is a good approximation of this superstring theory.
We choose the radius of $AdS_5$, $L_{AdS_5}$ to  be $1$, then the metric of $AdS_5\times \mathbf{RP}^5$ is
\be \label{metric} {\rm d}s^2={\rm d}s_{AdS_5}^2+{\rm d}s_{\mathbf{RP}^5}^2\,. \ee
The RR $5$-form fluxes is
\be \label{fiveform} F_5=4 (\omega_5+\tilde{\omega}_5)\,,\ee
where $\omega_5$ and $\tilde{\omega}_5$ are the volume forms on $AdS_5$ and $\mathbf{RP}^5$ with unit radius, respectively.

From $L_{AdS_5}=1$, one can get that~\cite{Giombi:2020kvo} in the large $N$ limit,
\be 4\pi g_s N \alpha^\prime=1\,,\ee
which leads to
\be \alpha^\prime=\sqrt{\frac2{\lambda}}\,,\label{lambda}\ee
by using the relation $g_{\rm YM}^2=8\pi g_s$ in the $SO(N)$ case
 \cite{Giombi:2020kvo}
 and the definition of the 't~Hooft coupling $\lambda\equiv g^2_{\rm YM}N$.

The discrete torsion for the Neveu-Schwarz $2$-form $B_{NS}$ and the RR $2$-form $B_{RR}$ are defined through
\begin{align}
  e^{2\pi  i\theta_{NS}} &\equiv \exp\left(i\int_{\mathbf{RP}^2}B_{NS}\right)=\pm 1\,,
\\
  e^{2\pi i \theta_{RR}} &\equiv \exp\left(i\int_{\mathbf{RP}^2}B_{RR}\right)=\pm 1\,.
\end{align}
where we pick up a $\mathbf{RP}^2$ inside $\mathbf{RP}^5$. When $(\theta_{NS}, \theta_{RR})=(0, 0)$ the gauge group of the dual theory is $SO(2n)$.
When $(\theta_{NS}, \theta_{RR})=(0, \frac12)$ the gauge group of the dual theory is $SO(2n+1)$.

\section{Chiral primary operators and the corresponding supergravity modes}\label{Sec:3}
We plan to compute the correlation functions of half-BPS chiral primary operators (CPOs) and various loop operators. These CPOs
 are constructed using the  six scalar fields $\Phi^i, i=1, \cdots, 6$,  which are in the adjoint representation of $SO(N)$ and the vector representation of $SO(6)_R$, the R-symmetry group of this theory.
Such CPOs are
\be \mathcal{O}^I=C^I_{i_1\cdots i_l}{\mathrm Tr}_\Box (\Phi^{i_1}\cdots \Phi^{i_l})\,, \ee
with $l\ge 2$. Here the trace is taken in the fundamental representation of $SO(N)$ and $C^I$ is in the tracelass $l$-th totally symmetric representation of $SO(6)_R$.
 We choose $C^I$ to satisfy
 \be  C_{i_1\cdots i_l}^I C^{J i_1\cdots i_l}=\delta^{IJ}\,,\ee
 here $C^{Ji_1\cdots i_l}$ is defined as $C^{Ji_1\cdots i_l}=\delta^{i_1j_1}\cdots \delta^{i_l j_l}C^J_{j_1\cdots j_l}$.
 Since $\Phi^i$'s are  $N\times N$ anti-symmetric matrices, $l$ should be even for $\mathcal{O}^I$ being non-vanishing.
This constraint is new compared to the case where the gauge group is $SU(N)$.

When $l\ll N$, the holographic description of $\mathcal{O}^I$ is expressed in terms of fluctuations of the background fields in IIB supergravity
on $AdS_5\times \mathbf{RP}^5$, \footnote{We use the notation that $m, n, \cdots $ refer to the coordinates in $AdS_5\times \mathbf{RP}^5$, $\mu, \nu, \cdots$ refer to the
ones in the $AdS_5$ part and $\alpha, \beta, \cdots$ refer to the ones in the $\mathbf{RP}^5$ part. The underlined indices refer to the target space ones. }
\begin{gather} G_{\underline{mn}}=g_{\underline{mn}}+h_{\underline{mn}}\,,\\
F_{\underline{m_1\cdots{m}_5}}=f_{\underline{m_1 \cdots m_5}}+\delta f_{\underline{m_1 \cdots m_5}}\,,\qquad  \delta f_{\underline{m_1 \cdots m_5}}=5 \nabla_{[\underline{m}_1}a_{\underline{m_2\cdots m_5}]}\,,
\end{gather}
where $g_{\underline{mn}}$ and $f_{\underline{m_1\cdots m_5}}$ are the background fields \eqref{metric} and \eqref{fiveform},
 $h_{\underline{mn}}$ and $\delta f_{\underline{m_1\cdots m_5}}$ are fluctuations.

The fluctuations dual to half-BPS CPOs are \cite{Lee:1998bxa}
\begin{align}
h_{\underline{\mu\nu}} &=-\frac{6}{5}\Delta s g_{\underline{\mu\nu}}+\frac{4}{\Delta+1}\nabla_{(\underline{\mu}}\nabla_{\underline{\nu})}s\,,\label{hmunu}
\\
h_{\underline{\alpha\beta}} &=2 \Delta s g_{\underline{\alpha\beta}}\,,
\\
a_{\underline{\mu_1\cdots\mu_4}} &=-4 \epsilon_{\underline{\mu_1\cdots \mu_5}}\nabla^{\underline{\mu_5}} s\,,\label{a4}
\\
a_{\underline{\alpha_1\cdots \alpha_4}} &=4 \sum_I \epsilon_{\underline{\alpha\alpha_1\cdots \alpha_4}} s^I(x)\nabla^{\underline{\alpha}}Y^I(y)\,.
\end{align}
Here $s(x, y)=\sum_I s^I(x)Y^I(y)$ with $x, y$ being coordinates in $AdS_5$ part and $\mathbf{RP}^5$ part, respectively.
$(\underline{\mu\nu})$ in \eqref{hmunu} means to take the traceless symmetric part.
$Y^I(y)$ is the ``scalar spherical harmonics'' on $\mathbf{RP}^5$ satisfying, \be \nabla^{\underline{\alpha}}\nabla_{\underline{\alpha}}Y^I=-\Delta (\Delta+4) Y^I\,. \ee
They are in the $[0, \Delta, 0]$ representation of $SO(6)_R$ and we choose the normalization of $Y^I$ to be the same  as the one in~\cite{Lee:1998bxa}. Since $\mathbf{RP}^5=S^5/\mathbf{Z}_2$, locally $Y^I$ is the same as the scalar spherical harmonics on $S^5$. $\Delta$ is dual to the conformal dimension of the CPO. For the case at hand, we have $\Delta=l$ since it is protected by supersymmetry. Recall that $l$ should be even. In the supergravity side, this is from the fact that  the $\mathbf{Z}_2$ projection of the  fields on $AdS_5\times S^5$ gives the fields on     $AdS_5\times \mathbf{RP}^5$.
$\epsilon_{\underline{\mu_1\cdots \mu_5}}$ and $\epsilon_{\underline{\alpha_1\cdots \alpha_5}}$ are the anti-symmetric tensors
corresponding to the volume form of $AdS_5$ and $\mathbf{RP}^5$, respectively. The background five-form field strength can then be expressed as
\be F_{\underline{\mu_1\cdots \mu_5}} =-4\epsilon_{\underline{\mu_1\cdots \mu_5}}\,, \qquad
F_{\underline{\alpha_1\cdots \alpha_5}} =-4\epsilon_{\underline{\alpha_1\cdots \alpha_5}}\,.\ee

\section{OPE of Wilson loops in the fundamental representation\label{fudamental}}
We consider half-BPS Wilson loop in the $SO(N)$ theory in  Euclidean space $\mathbf{R}^4$,
\be W_\Box[C]=\frac{1}{N}  {\mathrm Tr}_\Box \mathcal{P} \exp \left[\oint_C i\Big(A_\mu(x) \dot{x}^\mu+ i |\dot{x}|\Theta_j \Phi^j(x)\Big){\rm d}s\right], \ee
where the contour $C$ is $x^\mu(s)=(a \cos s, a\sin s, 0, 0)$, $\dot{x}^\mu=\frac{\partial x^\mu}{\partial s}$,
and $\Theta^j$ is a constant unit $6$-vector. The trace is taken in the fundamental representation.
For the dual description, we use the Euclidean $AdS_5$ ($EAdS_5$) in the Poincar\`e coordinates, such that the metric is
\be {\rm d}s^2=\frac{1}{z^2}({\rm d}z^2+{\rm d}x^{\underline{i}} {\rm d}x_{\underline{i}})\,. \ee
The action of the fundamental string (F-string) is
\be S=\frac{1}{2\pi\alpha^\prime} \int {\rm d}^2\sigma \sqrt{{\rm det}g_{\mu\nu}}\,, \ee
with the induced metric $g_{\mu\nu}$ being
\be g_{\mu\nu}=\frac{\partial x^{\underline{\rho}}}{\partial \sigma^\mu} \frac{\partial x^{\underline{\kappa}}}{\partial \sigma^\mu}g_{\underline{\rho\kappa}}\,.\ee

As for the F-string solution dual to the circular Wilson loop, we choose the worldsheet coordinates
to be $(z, s)$.
The corresponding classical F-string solution can be parameterized as~\cite{Berenstein:1998ij, Drukker:1999zq}
\be x^1=\sqrt{a^2-z^2}\cos s\,,\qquad x^2=\sqrt{a^2-z^2}\sin s\,,\qquad x^3=x^4=0\,. \ee
The worldsheet of this F-string has the topology of $EAdS_2$ and is entirely embedded within the $EAdS_5$ region of the background geometry.\footnote{In the following, we will sometimes use AdS to refer to EAdS for simplicity. It is expected that this will not result in any confusion.}

Taking into account the boundary terms from Legendre transformation \cite{Drukker:1999zq}, the on-shell action of this F-string is given by~\cite{Berenstein:1998ij, Drukker:1999zq}
\be  S_{F1}=\frac{1}{2\pi \alpha^\prime} (-2\pi)=-\frac1{\alpha^\prime}\,.\ee
Using \eqref{lambda}, we get \cite{Giombi:2020kvo}
\be S_{F1}=-\sqrt{\frac{\lambda}{2}}\,.\ee
Thus the holographic prediction for the {\it vev} of the Wilson loop
is
\be \langle W_\Box[C] \rangle= \exp\sqrt{\frac{\lambda}{2}}\,,\ee
in the large $N$ and large $\lambda$ limit.

When probing $W_\Box[C] $ from a distance $L$ much larger than its radius $a$. The operator product expansion (OPE) of $W_\Box [C]$ is
\be W_\Box [C]=\langle W_\Box[C]\rangle\left(1+\sum_{i, n}C_i^n a^{\Delta_i^n} O_i^n\right)\,, \label{OPE} \ee
where $\Delta_i^n$ are the conformal weight of the operator $O_i^n$. $O_i^0$ is the $i$-th primary field and $O_i^n$'s with $n>0$ are its conformal descends.

To extract the OPE coefficients of the half-BPS CPOs $O^I$ with normalized two-point functions, we can compute the normalized correlation of this Wilson loop and the half-BPS CPO $\mathcal{O}^I$,\footnote{$\boldsymbol{x}$ is the coordinate in $\mathbf{R}^4$.}
\be \langle \langle \mathcal{O}^I(\boldsymbol{x}) \rangle\rangle_{W_\Box[C]}\equiv\frac{\langle W_\Box[C]\mathcal{O}^I(\boldsymbol{x}) \rangle}{\sqrt{{\mathcal{N}_{\mathcal{O}^I}}}\langle W_\Box[C] \rangle }\,, \ee
where  $\mathcal{N}_{\mathcal{O}^I}$ is defined by the two point function of $\mathcal{O}^I$'s,
\be \langle \mathcal{O}^I(\mathbf{y})\mathcal{O}^J(\mathbf{z}) \rangle=\frac{\delta^{IJ}\mathcal{N}_{\mathcal{O}^I}}{|\mathbf{y}-\mathbf{z}|^{2\Delta_{\mathcal{O}^I}}}\,.  \ee
Take the OPE limit where $L=\sqrt{\boldsymbol{x}^2}\gg a$, we have
\be \langle \langle \mathcal{O}^I(\boldsymbol{x}) \rangle\rangle_{W_\Box[C]}=C_{\Box, \mathcal{O}}\frac{a^\Delta}{L^{2\Delta}}\,. \ee
The goal is to compute $C_{\Box, \mathcal{O}}$ holographically, which is the OPE coefficient of the primary operator $\mathcal{O}^I$ in the expansion~(\ref{OPE}).

For this goal, we need to calculate the change of F-string action due to the fluctuations of the background fields dual to $\mathcal{O}^I$ \cite{Berenstein:1998ij},
\be \delta S_{F1}=\frac{1}{2\pi \alpha^\prime} \int d^2 \sigma \sqrt{{\rm det}g_{\mu\nu}}\,\frac{1}{2} g^{\mu\nu} \frac{\partial x^{\underline{\rho}}}{\partial \sigma^\mu}
\frac{\partial x^{\underline{\kappa}}}{\partial \sigma^\nu}h_{\underline{\rho\kappa}}\,,\ee
where $\sigma^\mu$'s are worldsheet coordinates and $x^{\underline{\rho}}=x^{\underline{\rho}}(\sigma^\mu)$ expresses how the string worldsheet is embedded in the spacetime.

Then we write $s^I$ as $s^I({\boldsymbol{x}, z})=\int d^4\boldsymbol{x}^\prime G_\Delta(\boldsymbol{x}^\prime; \boldsymbol{x}, z) s_0^I(\boldsymbol{x}^\prime)$, here $s_0^I$
is a source for $\mathcal{O}^I$ on the boundary, and
\be G_\Delta(\boldsymbol{x}^\prime; \boldsymbol{x}, z)=c \left(\frac{z}{z^2+|\boldsymbol{x}-\boldsymbol{x}^\prime|^2}\right)^\Delta\,,\label{propagator}\ee
is the boundary-to-bulk propagator with the constant $c$ being \footnote{This constant $c$ is $\sqrt{2}$ times the corresponding constant in the $AdS_5\times S^5$ case given in~\cite{Freedman:1998tz, Lee:1998bxa}, due to the fact that $\mathbf{RP}^5=S^5/\mathbf{Z}_2$. For more details, see Appendix~\ref{appendix}.}
\be c=\frac{\Delta+1}{2^{(3-\Delta)/2} N\sqrt{\Delta}}\,.\label{constant} \ee
Then the correlation function is given by
\be \langle \langle \mathcal{O}^I(\boldsymbol{x}) \rangle \rangle_{W_\Box[C]}=-\frac{\delta S_{F1}}{\delta s_0^I(\boldsymbol{x})}\bigg|_{s_0^I=0}\,. \ee

In the OPE limit, we have
\be  G_\Delta(\boldsymbol{x}^\prime, \boldsymbol{x}, z)\simeq c\frac{z^\Delta}{L^{2\Delta}},\ee
\be \partial_{\underline{\mu}}s^I\simeq \delta^{z}_{\underline{\mu}}\frac{\Delta}{z}s^I\,,\ee
\be \partial_{\underline{\mu}}\partial_{\underline{\nu}}s^I\simeq \delta^{z}_{\underline{\mu}}\delta^z_{\underline{\nu}}\frac{\Delta(\Delta-1)}{z^2}s^I\,.\ee
Using these and the fact that in the Poincar\`e coordinates
\be \Gamma^z_{\underline{\mu\nu}}=z g_{\underline{\mu\nu}}-\frac{2}{z} \delta^z_{\underline{\mu}}\delta^z_{\underline{\nu}}\,. \ee
Then from \eqref{hmunu}, we get
\be h_{\underline{\mu\nu}}\simeq -2 \Delta  g_{\underline{\mu\nu}}s^I Y^I+\frac{4\Delta}{z^2}\delta^z_{\underline{\mu}}\delta^z_{\underline{\nu}}s^I Y^I\,,\label{newh} \ee

The induce metric is
\be g_{ss}=\frac{a^2-z^2}{z^2}\,,\qquad g_{sz}=0\,,\qquad g_{zz}=\frac{a^2}{z^2(a^2-z^2)}\,. \ee
We have \be{\rm det}(g_{\mu\nu})=\frac{a^2}{z^4}\,. \ee
From these, we obtain
\be g^{\mu\nu}\frac{\partial x^{\underline{\rho}}}{\partial \sigma^\mu} \frac{\partial x^{\underline{\rho}}}{\partial \sigma^\nu} h_{\underline{\rho\kappa}}=-2
\Delta  \frac{z^2}{a^2} s^IY^I\,. \ee
Then the variation of the F-string action is
\be \delta S_{F1}=-\frac{\Delta Y^I}{\pi \alpha^\prime a} \int d^2\sigma s^I \,.\ee
Using \eqref{propagator}, we get
\begin{align}  \langle \langle \mathcal{O}^I(\boldsymbol{x}) \rangle \rangle_{W_\Box[C]}&=-\frac{\delta S_{F1}}{\delta s_0^I(\boldsymbol{x})}\bigg|_{s_0^I=0}\nonumber\\
&=\frac{\Delta Y^I(y)c}{\pi \alpha^\prime a L^{2\Delta}}\int {\rm d}^2\sigma z^\Delta=\frac{\Delta Y^I(y)c}{\pi \alpha^\prime a L^{2\Delta}}\int_0^\pi {\rm d}\psi \int_0^a {\rm d}z z^\Delta\nonumber\\
&=Y^I(y)\frac{2c\Delta}{\alpha^\prime (\Delta+1)}\frac{a^\Delta}{L^{2\Delta}}\,.
\end{align}
Now by using \eqref{lambda} and  \eqref{constant}, we obtain
\be \langle \langle \mathcal{O}^I(\boldsymbol{x}) \rangle \rangle_{W_\Box[C]}=Y^I(y)2^{\Delta/2-1}\frac{\sqrt{\lambda \Delta}}{N}\frac{a^\Delta}{L^{2\Delta}}\,. \ee
Thus the OPE coefficient is\footnote{Here $y$ is the image of $\Theta$ under the map $S^5\to \mathbf{RP}^5$. 
This also applies for the case of the D3 brane in the next subsection.}
\be\mathcal{C}_{\Box, \mathcal{O}}=Y^I(y)2^{\Delta/2-1}\frac{\sqrt{\lambda \Delta}}{N}\,. \ee
We use the convention that the factor $Y^I(y)$ is not included in the OPE coefficient which leads to
\be C_{\Box, \mathcal{O}}=2^{\Delta/2-1}\frac{\sqrt{\lambda \Delta}}{N}\,.\label{cbox}\ee
The above result expressed in terms of $\lambda, N$ and $\Delta$, 
is identical to the result obtained in the $SU(N)$ case~\cite{Berenstein:1998ij}.
Since the string worldsheet is an $AdS_2$ subspace completely embedded inside the $AdS_5$ part of the background geometry, the change from $S^5$ to $\mathbf{RP}^5$ does not impact the calculation of the coupling between the supergravity modes and the string worldsheet.
The relation between $\alpha^\prime$ and $\lambda$ in the $SO(N)$ case is $\alpha^\prime=\sqrt{2/\lambda}$, which has an extra factor of $\sqrt{2}$, compared with the relation $\alpha^\prime=\sqrt{\lambda}$ in the $SU(N)$ case. While the coefficient inside the bulk-to-boundary propagator, $c$, is also changed as $c_{\rm SO}=\sqrt{2}c_{\rm SU}$. These two effects cancel with each other, thus the results of the OPE coefficients in terms of $\lambda, N, \Delta$ are identical in the cases of both $SO(N)$ and $SU(N)$.
But one should keep in mind that $\Delta$ should be even for the case of $SO(N)$.

\section{OPE of  WH loops in the symmetric representation\label{symmetric}}
In this section, we will compute the OPE coefficients of the half-BPS circular Wilson loop-'t~Hooft loops in the symmetric representation.
The WH loop comes from the worldline of a dyon which carry both electric and magnetic charges of the gauge theory.
In this section, we will only consider the case when the dyon is in the $n$-th symmetric representation of $g$ and the $m$-th symmetric
representation of $^Lg$.\footnote{A more precise description of such WH loops was provided in the Section~\ref{introduction}.}
 When $m=0$, we get the following Wilson loops in the $n$-th symmetric representation,
\be W_{S_n}[C]=\frac{1}{{\rm dim} S_n}  {\rm Tr}_{S_n} \mathcal{P} \exp \left[\oint_C i\Big(A_\mu(x) \dot{x}^\mu+i|\dot{x}|\Theta_j \Phi^j(x)\Big){\rm d}s\right]\,, \ee
where $S_n$ denotes the $n$-th symmetric representation of $SO(N)$ and ${\rm dim }S_n$ denotes its dimension.

Non-trivially generalizing the results in \cite{Drukker:2005kx}, it was proposed in \cite{Chen:2006iu} that for $SU(N)$ case,
the Wilson-'t Hooft (WH) loop is dual to D3-branes in $AdS_5\times S^5$. In \cite{Giombi:2020kvo}, D3-brane dual to Wilson loop in the symmetric representation
for the $SO(N)$ case was given. We expect the generalization of the solution in \cite{Chen:2006iu} to $AdS_5\times \mathbf{RP}^5$ case will provide
the dual description of the WH loop in the symmetric representation for the $SO(N)$ case.

We start with the coordinate system in $EAdS_5$ such that the metric take the form \cite{Drukker:2005kx}
\be {\rm d}s^2=\frac{1}{z^2}({\rm d}z^2+{\rm d}r_1^2+r_1^2{\rm d}\psi^2+{\rm d}r_2^2+r_2^2{\rm d}\phi^2)\,.\label{poincare}\ee
The boundary of $EAdS_5$ is now at $r\to\infty$ and $\eta=0$.
  In this  coordinate the $AdS_5$ part of the RR $4$-form potential is chosen to be
  \be C^{AdS}_4=\frac{r_1r_2}{z^4}\,{\rm d}r_1 \wedge {\rm d}\psi \wedge {\rm d}r_2 \wedge {\rm d}\phi\,. \ee
We place the WH loop on the boundary at $r_1=a, r_2=0$.
We make the following coordinate transformation,
\begin{align}
r_1&=\frac{a \cos\eta}{\cosh\rho-\sinh\rho \cos\theta}\,,\label{r1}\\
r_2&=\frac{a \sinh\rho \sin\theta}{\cosh\rho-\sinh\rho\cos\theta}\,,\label{r2}\\
z&=\frac{ a \sin\eta}{\cosh\rho-\sinh\rho\cos\theta}\,.\label{z}
\end{align}
The metric on $EAdS_5$ in this coordinate system is
\be {\rm d}s^2=\frac{1}{\sin^2\eta}\big[{\rm d}\eta^2+\cos^2\eta {\rm d}\psi^2+\sinh^2\rho({\rm d}\theta^2+\sin^2\theta {\rm d}\phi^2)\big]\,. \ee

We only  consider the case when  the theta angle in the field theory is zero. This corresponds to set the background RR zero form
potential (the axion), $C_0$, to be zero.
Then the action of the D3-brane in $AdS_5\times \mathbf{RP}^5$ background is
\be S^{\rm D3}=S^{\rm D3}_{\rm DBI}+S^{\rm D3}_{\rm WZ}\,, \ee
where
\begin{align}
S^{\rm D3}_{\rm DBI}&=T_{\rm D3}\int d^4\sigma\sqrt{{\rm det} (g+2\pi \alpha^\prime F)}\,,\\
S^{\rm D3}_{\rm WZ}&=-T_{\rm D3}\int P[C_4]\,.
\end{align}
Here $g$ is the induced metric on the D3-brane, $F$ the electromagnetic field on the D3-brane worldvolume, $P[C_4]$ is the pull-back of $C_4$ to the worldvolume and the D3-brane tension reads
\begin{equation}
T_{\rm D3}=\frac{1}{(2\pi)^3\alpha^{\prime2}g_s}=\frac{N}{2\pi^2}\,,
\end{equation}
where the relations $\alpha^\prime=\sqrt{2/\lambda}$ and $g_{\rm YM}^2=8\pi g_s$ in $SO(N)$ case have been used.

For the D3-brane dual to the above WH loop, we take the worldvolume coordinates to be $\rho, \psi, \theta, \phi$, and $\eta=\eta(\rho)$
on the worldvolume. We also need to turn on the components $F_{\psi\rho}$ and $F_{\theta\phi}$ of the electromagnetic field  strength on the D3-brane worldvolume.

The D3-brane solution, obtained by adjusting the solution in~\cite{Chen:2006iu} to the $SO(N)$ case, is given by
\begin{align}
\sin\eta&=\kappa^{-1}\sinh \rho,\qquad \kappa=\sqrt{\frac{n^2\lambda}{32 N^2}+\frac{2\pi^2 m^2}{\lambda}}\,,\\
F_{\psi\rho}&=\frac{in\lambda}{16\pi N\sinh^2\rho}\,,\qquad F_{\theta\phi}=\frac{m \sin\theta}{2}\,.
\end{align}
Let us introduce dual 't~Hooft coupling\footnote{This results is from $\tilde{\lambda}=^Lg^2_{{\rm YM}}N $, with the dual Yang-Mills coupling $^Lg_{{\rm YM}}=\frac{4\pi \sqrt{n_g}}{g_{{\rm YM}}}$ when $\theta_{{\rm YM}}=0$ \cite{Dorey:1996hx, Vafa:1997mh,  Argyres:2006qr}.}, $\tilde{\lambda}=\frac{16\pi^2N^2 n_g}{\lambda}$,
where $n_g=1$ for $g=spin(2n)$, and $n_g=2$ for $g=spin(2n+1)$.
Then we can express $\kappa$ as
\be \kappa=\frac{1}{4N}\sqrt{\frac{n^2\lambda}{2}+\frac{2m^2\tilde{\lambda}^2}{n_g}}\,.\ee
Taking into account the boundary terms, the on-shell action of the D3-brane is
\be S^{\rm D3}=-2 N (\kappa \sqrt{1+\kappa^2}+\sinh^{-1}\kappa)\,. \ee
Thus the holographic prediction of vacuum expectation value of WH is
\be \langle {WH}_{S_n, S_m}[C]\rangle=\exp\big[2N (\kappa \sqrt{1+\kappa^2}+\sinh^{-1}\kappa)\big]\,. \ee
When we take $m=0$, this D3-brane solution is the same as the one in \cite{Giombi:2020kvo}, though in different
coordinates. Furthermore, the holographic prediction for $\langle W_{S_n}\rangle$  is consistent with the results from localization \cite{Giombi:2020kvo} in the large $\lambda$ limit
with $\kappa$ fixed.

Now we holographically  compute the correlator of ${WH}_{S_n, S_m[C]}$ and $\mathcal{O}^I(\mathbf{x})$
\be \langle \langle \mathcal{O}^I(\mathbf{x})\rangle\rangle_{{WH}_{S_n, S_m}[C]}\equiv \frac{\langle {WH}_{S_n, S_m}[C]\mathcal{O}^I(\mathbf{x}) \rangle}{\sqrt{\mathcal{N}_{\mathcal{O}}}\langle {WH}_{S_n, S_m}[C] \rangle}\,, \ee in the OPE limit $L\gg a$, and  extract the OPE coefficient $C_{WH_{S_n, S_m}, \mathcal{O}}$.
The change of the $S_{\rm DBI}^{\rm D3}$ due to the fluctuations of the background field is
\be \delta S_{\rm DBI}^{\rm D3}=T_{\rm D3}\int {\rm d}^4\sigma \sqrt{{\rm det}\mathcal{M}} \,\frac12 (\mathcal{M}^{-1})^{\mu\nu} \frac{\partial x^{\underline{\rho}}}{\partial \sigma^\mu}
\frac{\partial x^{\underline{\kappa}}}{\partial \sigma^\nu} h_{\underline{\rho}\underline{\kappa}}\,, \ee
where we have defined the matrix $\mathcal{M}=g+2\pi \alpha^\prime F$, $\sigma^\mu$'s are worldvolume coordinates.
By using  the result of  $h_{\underline{\rho \kappa}} $ in the OPE limit given in \eqref{newh} and  the above D3-brane solution, we obtain
\be \delta S_{\rm DBI}^{\rm D3}=4N \kappa^2 Y^I \int {\rm d}\rho {\rm d}\theta\, \frac{\sin \theta}{\sinh^2\rho}\left(-1-2\kappa^2+\frac{1-\sinh^2 \rho(\kappa^{-2}-\sin^2\theta)}{(\cosh\rho-\sinh\rho\cos\theta)^2}\right)s^I\,.\ee
Now we turn to compute the change of $S_{\rm WZ}$ due to the fluctuations of the background fields,
\be  \delta S_{\rm WZ}^{\rm D3}=-T_{\rm D3}\int  P[a_4].\ee
From \eqref{a4}, we have
\be a^I_{\underline{\mu \cdots \mu_4}}=-4\Delta z \epsilon_{\underline{\mu \cdots \mu_4} z} s^IY^I=-4 \Delta C_{\underline{\mu \cdots \mu_4}}s^IY^I. \ee
Thus \be \delta S_{\rm WZ}^{\rm D3}=4T_{\rm D3}\Delta Y^I \int P[C_4]s^I\,. \ee
From the coordinate transformation \eqref{r1}-\eqref{z}, we obtain
\be \delta S_{\rm WZ}^{\rm D3}=8N\Delta \kappa^4 Y^I\int {\rm d}\rho {\rm d}\theta \,\frac{\sin\theta}{\sinh^2\rho}\left(1-\frac{1}{\kappa^2}\frac{\sinh\rho \cos\theta}{\cosh\rho-\sinh\rho\cos\theta}\right)\,. \ee
Then the total change of the action is
\be\delta S^{\rm D3} =\delta S_{\rm DBI}^{\rm D3}+\delta S_{\rm WZ}^{\rm D3}=-4N\Delta Y^I\int_0^{\sinh^{-1}\kappa}  {\rm d}\rho \int_{0}^{\pi}{\rm d}\theta\frac{\sin\theta}{(\cosh\rho-\sinh\rho\cos\theta)^2}s^I\,. \ee

Using $s^I({\boldsymbol{x}, z})=\int {\rm d}^4\boldsymbol{x}^\prime G_\Delta(\boldsymbol{x}^\prime; \boldsymbol{x}, z) s_0^I(\boldsymbol{x}^\prime)$, we can compute $\langle \langle \mathcal{O}(\mathbf{x})\rangle\rangle_{{\rm WH}_{S_n, S_m }[C]}$ as
\be\langle \langle \mathcal{O}(\mathbf{x})\rangle\rangle_{WH_{S_n, S_m }[C]}=-\frac{\delta S^{\rm D3}}{\delta s^0(\mathbf{x})}\,. \ee
In the OPE limit, we have
\be \langle \langle \mathcal{O}(\mathbf{x})\rangle\rangle_{WH_{S_n, S_m }[C]}=\frac{a^\Delta}{L^{2\Delta}} \frac{4N\Delta c Y^I(y)}{\kappa^\Delta}\int_0^{\sinh^{-1}\rho}d\rho\int_0^\pi d\theta\frac{\sinh^\Delta\rho\sin\theta}{(\cosh\rho-\sinh\rho\cos\theta)^{2+\Delta}}\,.\ee
Performing the two integral we get
\be\mathcal{C}_{WH_{S_n, S_m}, \mathcal{O}}=\frac{2^{(\Delta+3)/2}}{\sqrt{\Delta}}Y^I(y)\sinh(\Delta \sinh^{-1}\kappa)\,. \ee
Thus\be C_{WH_{S_n, S_m}, \mathcal{O}}=\frac{2^{(\Delta+3)/2}}{\sqrt{\Delta}}\sinh(\Delta \sinh^{-1}\kappa)=\frac{i(-1)^{\Delta/2}2^{(\Delta+3)/2}}{\sqrt{\Delta}}V_\Delta(i\kappa)\,.\label{symmetric} \ee
Here  $V_n(x)=\sin   (n \cos^{-1}x)$ is one type             of the Chebyshev polynomials and we have use the fact that $\Delta$ is even.

The result for Wilson loop ($m=0$) in terms of $\kappa$ is $\sqrt{2}$ times the results in~\cite{Giombi:2006de} for $SU(N)$ case due to the change of $c$.\footnote{There is a sign typo in~\cite{Giombi:2006de} when the result was finally expressed using the Chebyshev polynomials.} 
Here we provide a brief explanation on this point.
Since the worldvolume of D3-brane is completely inside $AdS_5$, 
the calculations of the coupling between the supergravity modes and the D3-brane worldvolume
for both $SU(N)$ and $SO(N)$ cases are the same.
For the $SO(N)$ case, the relation between $\alpha^\prime$ and $\lambda$ reads
$\alpha^\prime=\sqrt{2/\lambda}$, 
while the relation $g_{\rm YM}^2=8\pi g_s$ in the $SO(N)$ case is also changed compared with the $SU(N)$ case.
However, their effects on $T_{\rm D3}$ cancel with each other. The relation between $T_{\rm D3}$ and $N$, 
i.e., $T_{\rm D3}=N/(2\pi^2)$ is unchanged. Formally, when we express the results in terms of $\kappa$ and $\Delta$, the only change is from the coefficient of the bulk-to-boundary propagator $c_{\rm SO}=\sqrt{2}c_{\rm SU}$. This leads to the above conclusion about the OPE coefficients.  However,
the relation between $\kappa$ and $\lambda$ is changed for the case of $SO(N)$, which is 
\be \kappa=\frac{n}{4N} \sqrt{\frac{\lambda}{2}}\,, \ee
while for the Wilson loop in the $n$-th symmetric representation of $spin(N)$ in the $SU(N)$ case,
the relation reads
 \be \kappa=\frac{n\sqrt{\lambda}}{4N}\,. \ee
Hence, 
the result in terms of $\lambda$ and $\Delta$ for the $SO(N)$ case is {\it not} just a constant multiplying the result in the $SU(N)$ case. 

Finally, to compare with the result about in $C_{\Box, \mathcal{O}}$ in \eqref{cbox}, we set $m=0$ in~(\ref{symmetric}) and take the $\kappa\to 0$ limit.
 Using $\kappa=\frac{n}{4N}\sqrt{\lambda/2}$ in this case, we obtain
\be  C_{W_{S_n}, \mathcal{O}}\simeq 2^{(\Delta+3)/2}\sqrt{\Delta}\kappa=2^{\Delta/2-1}\frac{\sqrt{\Delta \lambda}}{N}n,\ee
which is just $nC_{\Box, \mathcal{O}}$, as expected.

\section{OPE of Wilson loops in the anti-symmetric representation\label{antisymmetric}}

Let us consider the half-BPS circular Wilson loops in the rank-$k$ anti-symmetric representation of the gauge group $SO(N)$,
\be W_{A_k}[C]=\frac{1}{{\rm dim} A_k}  {\rm Tr}_{A_k} \mathcal{P} \exp \left[\oint_C i\Big(A_\mu(x) \dot{x}^\mu+i|\dot{x}|\Theta_j \Phi^j(x)\Big){\rm d}s\right]. \label{ak}\ee
They have a bulk description in terms of D5-brane with  $k$ units of fundamental string charge. The worldvolume of this D5-brane has topology $AdS_{2}\times S^{4}$. The D5 description of Wilson loop is valid in the large $N$ large $\lambda$ limit with $k/N$ fixed.

We can parameterize the unit $S^5$, $\sum_{i=1}^6z_i^2=1$ as
\begin{equation}\label{fivesphere}
    z_1=\cos\theta,\qquad z_{j+1}=\sin\theta w_j, \, j=1, \cdots, 5,
\end{equation}
with $\sum_{j=1}^5w_j^2=1$.
Then the metric of unit $S^5$ can be written as
\be d\Omega^2_5=d\theta^2+\sin^2\theta d\Omega^2_4\,, \ee
with $d\Omega^2_4$ the metric of unit $S^4$.

$\mathbf{RP}^5$ can be obtained from $S^5$ by identifying antipodal points $z_i\sim -z_i$. One way
to realize this is to view $\mathbf{RP}^5$ as the upper hemisphere of $S^5$ ($0\leq \theta\leq \pi/2$) with antipodal points on the equator ($\theta=\pi/2$) identified.
The metric of $\mathbf{RP}^5$ is thus given by
\be ds_{\mathbf{RP}^5}^2=d\theta^2+\sin^2\theta ds^{\prime 2}_4\,,  \qquad 0\leq \theta\leq \pi/2 \ee
where $ds^{\prime 2}_4=d\Omega_4^2$ when $\theta<\pi/2$ and the $ds^{\prime 2}_4$ is the metric of $\mathbf{RP}^4$ when $\theta=\pi/2$.

Hence, the metric of $AdS_{5}\times\mathbf{RP}^{5}$ reads
\begin{equation}
{\rm d}s^{2}=\cosh^{2}u({\rm d}\zeta^{2}+\sinh^{2}\zeta{\rm d}\psi^{2})+{\rm d}u^{2}+\sinh^{2}u({\rm d}\vartheta^{2}+\sin^{2}\vartheta{\rm d}\phi^{2})+{\rm d}\theta^{2}+\sin^{2}\theta{\rm d}s^{\prime 2}_{4}\,,\label{metricadsrp}
\end{equation}
with the radius of $AdS_{5}$ and $\mathbf{RP}^{5}$ set to be $1$. The
$AdS_{5}$ part of the above metric is written in the form of an $AdS_{2}\times S^{2}$
fibration for computation convenience and these coordinates are related
to the one in \eqref{poincare} by the following coordinate transformation,
\begin{align}
r_{1} & =\frac{a\cosh u\sinh\zeta}{\cosh u\cosh\zeta-\cos\vartheta\sinh u}\,,\\
r_{2} & =\frac{a\sinh u\sin\vartheta}{\cosh u\cosh\zeta-\cos\vartheta\sinh u}\,,\\
z & =\frac{a}{\cosh u\cosh\zeta-\cos\vartheta\sinh u}\,,\label{eq:D5z}
\end{align}
where $a$ is the radius of the Wilson loop.

With an $SO(6)_R$ transformation, we can set the $\Theta^I$ in \eqref{ak} to be $\Theta^I=(1, 0, \cdots, 0)$.  Then in $AdS_{5}\times\mathbf{RP}^{5}$,
the D5-brane dual to the this antisymmetric Wilson loop occupies the $AdS_{2}$
in the above metric with $u=0$, and wraps an $S^{4}$ submanifold
of $\mathbf{RP}^{5}$ at a constant polar angle $\theta_{k}$ (on
the upper hemisphere of $S^{5}$)~\cite{Giombi:2020kvo}.
The D5-brane worldvolume is the $AdS_{2}\times S^{4}\subset AdS_5\times \mathbf{RP}^{5}$  and
its metric reads
\begin{equation}
{\rm d}\tilde{s}^{2}={\rm d}\zeta^{2}+\sinh^{2}\zeta{\rm d}\psi^{2}+\sin^{2}\theta_k{\rm d}\Omega_{4}^{2}\,,.
\end{equation}
Turning on the worldvolume $U(1)$ gauge field $F_{\psi\zeta}$ to account for the $k$ units of fundamental brane charge, the action of the D5-brane in ${\rm AdS}_{5}\times\mathbf{RP}^{5}$
background can be written as
\begin{equation}
S^{{\rm D5}}=S_{{\rm DBI}}^{{\rm D5}}+S_{{\rm WZ}}^{{\rm D5}}\,,
\end{equation}
where
\begin{align}
S_{{\rm DBI}}^{{\rm D5}} & =T_{{\rm D5}}\int{\rm d}^{6}\sigma\sqrt{\det(g+2\pi\alpha^{\prime}F)},\\
S_{{\rm WZ}}^{{\rm D5}} & =-2\pi\alpha^{\prime}iT_{{\rm D5}}\int F\wedge P[C_{(4)}].
\end{align}
In the above equations the tension of D5-brane reads
\begin{equation}
T_{{\rm D}5}=\frac{1}{g_{s}(2\pi)^{5}(\alpha^{\prime})^{3}}=\frac{N}{8\pi^{4}}\sqrt{\frac{\lambda}{2}}\,,
\end{equation}
and the self-dual 4-form potential is chosen to be~\cite{Yamaguchi:2006tq}
\begin{equation}
C_{(4)}=4\Big(\frac{u}{8}-\frac{\sinh4u}{32}\Big){\rm d}H_{2}\wedge{\rm d}\Omega_{2}-\Big(\frac{3}{2}\theta-\sin2\theta+\frac{1}{8}\sin4\theta\Big){\rm d}\Omega_{4}\,.
\end{equation}
Here ${\rm d}H_2$ is the volume form of the unit $AdS_2$, $\sinh \zeta {\rm d}\zeta \wedge {\rm d}\psi$, ${\rm d}\Omega_2$ is the volume form of the unit $S^2$, $\sin \vartheta {\rm d} \vartheta \wedge d\phi $, and  ${\rm d}\Omega_4$ is the volume form of the unit $S^4$.

The fact that the flux of the worldvolume gauge field equal to $k$  together with the brane equations of motion  give rise to
the condition~\cite{Yamaguchi:2006tq}\footnote{Here we restrict $k<N/2$. Then we have $\theta_k<\pi/2$. It is proposed \cite{Giombi:2020kvo} that the D5 brane doubly wrapping $\mathbf{RP}^4$ at $\theta=\pi/2$ is dual to the antisymmetric Wilson loops with $k=N/2$ for even $N$. }
\begin{equation}
\theta_{k}-\sin\theta_{k}\cos\theta_{k}=\pi\frac{k}{N}\,,
\end{equation}
and
the worldvolume gauge field is
\begin{equation}
F_{\psi\zeta}=\frac{i\sqrt{\lambda/2}\sinh\zeta\cos\theta_{k}}{2\pi}\,.\label{Fpsizeta}
\end{equation}
The on-shell D5-brane  DBI and WZ action is
\begin{align}
S_{{\rm DBI}}^{{\rm D5}} &=\frac{2N}{3\pi}\sqrt{\frac{\lambda}{2}}\int{\rm d}\zeta\sinh\zeta\sin^{5}\theta_{k}\,.\\
S_{{\rm WZ}}^{{\rm D5}} & =\frac{4iN}{3}\int{\rm d}\zeta\:F_{\psi\zeta}\Big(\frac{3}{2}\theta-\sin2\theta+\frac{1}{8}\sin4\theta\Big)\,,
\end{align}
Adding appropriate boundary terms~\cite{Yamaguchi:2006tq}, the on-shell action for
the D5-brane is
\begin{equation}
S^{\rm D5}=S_{{\rm DBI}}^{\rm D5}+S_{WZ}^{\rm D5}+S_{\rm bdy}^{\rm D5}=-\frac{2N}{3\pi}\sqrt{\frac{\lambda}{2}}\sin^{3}\theta_{k}\,.
\end{equation}
Thus the holographic prediction for the expectation value of the Wilson loop in the rank $k$  antisymmetric representation
is given by
\begin{equation}
\langle W_{A_{k}}\rangle=\exp\Big(\frac{2N}{3\pi}\sqrt{\frac{\lambda}{2}}\sin^{3}\theta_{k}\Big)\,.
\end{equation}

The variation of the DBI part of the action to the first order in
the fluctuation $h_{\mu\nu}$ and $h_{\alpha\beta}$ is
\begin{align}
\delta S_{{\rm DBI}}^{{\rm D5}} & =T_{{\rm D5}}\int{\rm d}^{6}\sigma\sqrt{\det(g+2\pi\alpha^{\prime}F)}\left((g+2\pi\alpha^{\prime}F\right)^{-1})^{mn}\nonumber\\
&\times \frac{1}{2}(h_{\underline{\mu\nu}}\partial_{m}X^{\underline{\mu}}\partial_{n}X^{\underline{\nu}}+h_{\underline{\alpha\beta}}\partial_{m}X^{\underline{\alpha}}\partial_{n}X^{\underline{\beta}})\nonumber \\
 & =\frac{N}{3\pi}\sqrt{\frac{\lambda}{2}}\sin^{5}\theta_{k}\int{\rm d}\zeta\,\sinh\zeta\Big(\frac{-4\Delta}{\cosh^{2}\zeta\sin^{2}\theta_{k}}+8\Delta\Big)s^{\Delta}Y^{\Delta,0}(\theta_{k})\,,
\end{align}
where we have used the D5 solution $z=a/\cosh\zeta$, c.f., \eqref{eq:D5z}.
The variation of the WZ part of the action to the first order in the
fluctuation is given by\footnote{Here and in the following, we have used the fact that integrating over $S^4$ selects the $SO(5)$ invariant harmonics. Then the harmonics only depends on $\theta_k$.}
\begin{align}
\delta S_{{\rm WZ}}^{{\rm D5}} & =-2\pi\alpha^{\prime}iT_{{\rm D5}}\int F\wedge P[a_{(4)}]\nonumber \\
 & =-2\pi\alpha^{\prime}iT_{{\rm D5}}\int{\rm d}\psi{\rm d}\zeta{\rm d}\sigma_{1}{\rm d}\sigma_{2}{\rm d}\sigma_{3}{\rm d}\sigma_{4}\,\mu(\Omega_{4})\,F_{\psi\zeta}\times4\sin^{4}\theta S^{I}\partial_{\theta}Y^{I}\nonumber \\
 & =\frac{8N}{3\pi}\sqrt{\frac{\lambda}{2}}\cos\theta_{k}\sin^{4}\theta_{k}\int{\rm d}\zeta\,\sinh\zeta s^{\Delta}\partial_{\theta_{k}}Y^{\Delta,0}(\theta_{k})\,,
\end{align}
where 4-form fluctuation is given by
\begin{equation}
a_{\sigma_{1}\sigma_{2}\sigma_{3}\sigma_{4}}=4\sin^{4}\theta\,\mu(\Omega_{4})\sum s^{I}\partial_{\theta}Y^{I}
\end{equation}
with $\sigma_{1},\sigma_{2},\sigma_{3},\sigma_{4}$ the coordinates
on the $S^{4}$ and the corresponding measure $\mu(\Omega_{4})$ is \be \mu(\Omega_{4})=\sin^{3}\sigma_{1}\sin^{2}\sigma_{2}\sin\sigma_{3}.
\ee
Thus the variation of the D5 action to the first order is given by
\begin{equation}
\delta S^{{\rm D5}}=\delta S_{{\rm DBI}}^{{\rm D5}}+\delta S_{{\rm WZ}}^{{\rm D5}}\,.
\end{equation}
The normalized correlation function between the Wilson loop and the chiral primary
operator is evaluated as
\begin{equation}
\frac{\langle W_{A_k}(\mathcal{C})\mathcal{O}_{\Delta}(L)\rangle}{\sqrt{\mathcal{N}_{\mathcal{O}}}\langle W(\mathcal{C})\rangle}=-\frac{\delta S_{{\rm D5}}}{\delta s_{0}}
\end{equation}
Recall that
\begin{equation}
s^{I}(\vec{x},z)=\int{\rm d}^{4}\vec{x}^{\prime}G_{\Delta}(\vec{x}^{\prime},\vec{x},z)s_{0}^{I}(\vec{x})\,,
\end{equation}
where the bulk-to-boundary propagator
\begin{equation}
G_{\Delta}(\vec{x}^{\prime},\vec{x},z)=c\left(\frac{z}{z^{2}+|\vec{x}-\vec{x}^{\prime}|^{2}}\right)^{\Delta}\simeq c\frac{z^{\Delta}}{L^{2\Delta}}
\end{equation}
and the D5 solution $z=a/\cosh\zeta$. The only integral one needs
to perform is
\begin{equation}
\int_{0}^{\infty}{\rm d}\zeta\,\frac{\sinh\zeta}{\cosh^{\Delta+2}\zeta}=\frac{1}{\Delta+1}\,.
\end{equation}
Hence we obtain
\begin{equation}
\frac{\langle W_{A_{k}}\mathcal{O}_{\Delta}(L)\rangle}{\sqrt{\mathcal{N}_{\mathcal{O}}}\langle W_{A_{k}}\rangle}=\frac{a^{\Delta}}{L^{2\Delta}}\frac{2^{\Delta/2}\mathcal{N}_{\Delta}}{3\pi}\sqrt{\frac{\lambda}{\Delta}}\frac{\Delta+3}{\Delta-1}\sin^{3}\theta_{k}\big[2(\Delta+1)\cos\theta_{k}C_{\Delta-1}^{(2)}-\Delta C_{\Delta}^{(2)}\big]\,,
\end{equation}
where we have used the following results  about the  $SO(5)$ invariant harmonics~\cite{Giombi:2006de}\footnote{Here we use the normalization of spherical hamonic function in~\cite{Lee:1998bxa}, so the $\mathcal{N}_\Delta$ obtained here is different from the one in~\cite{Giombi:2006de}. This difference will disappear when we express the final result in terms of $Y^{\Delta, 0}(0)$.}
\begin{equation}
Y^{\Delta,0}(\theta_{k})=\mathcal{N}_{\Delta}C_{\Delta}^{(2)}(\cos\theta_{k})\,,
\end{equation}
and
\begin{equation}
\Delta Y^{\Delta, 0}(\theta_{k})+\frac{\cos\theta_{k}}{\sin\theta_{k}}\partial_{\theta_{k}}Y^{\Delta, 0}(\theta_{k})=\mathcal{N}_{\Delta}\Big[\frac{\Delta}{\sin^{2}\theta_{k}}C_{\Delta}^{(2)}-(\Delta+3)\frac{\cos\theta_{k}}{\sin^{2}\theta_{k}}C_{\Delta-1}^{(2)}\Big]\,.
\end{equation}
The normalization factor $\mathcal{N}_{\Delta}$ is obtained by
\begin{equation}
Y^{\Delta, 0}(0)=\mathcal{N}_{\Delta}C_{\Delta}^{(2)}(1)=\mathcal{N}_{\Delta}\frac{(\Delta+3)!}{6\Delta!}\,,
\end{equation}
and thus
\begin{equation}
\mathcal{N}_{\Delta}=\frac{6\Delta!}{(\Delta+3)!}Y^{\Delta, 0}(0)\,.
\end{equation}
We then obtain
\begin{equation}
\frac{\langle W_{A_{k}}\mathcal{O}_{\Delta}(L)\rangle}{\sqrt{\mathcal{N}_{\mathcal{O}}}\langle W_{A_{k}}\rangle}=\frac{a^{\Delta}}{L^{2\Delta}}Y^{\Delta,0}(0)\frac{2^{\Delta/2}}{3\pi}\sqrt{\Delta\lambda}\frac{6(\Delta-2)!}{(\Delta+1)!}\sin^{3}\theta_{k}\big[2(\Delta+1)\cos\theta_{k}C_{\Delta-1}^{(2)}-\Delta C_{\Delta}^{(2)}\big]\,.
\end{equation}
Using the recurrence relation~\cite{Higuchi:1986wu, szego}
\begin{equation}
\Delta C^{(\lambda)}_\Delta (x)=
2 (\Delta +\lambda -1) x C^{(\lambda)}_{\Delta -1}(x)
-(\Delta +2\lambda -2) C^{(\lambda)}_{\Delta -2}(x)\,,
\end{equation}
we finally arrive
\begin{equation}
C_{W_{A_k}, \mathcal{O}} = \frac{2^{\Delta/2}}{3\pi}\sqrt{\Delta \lambda} \sin^3\theta_k
\frac{6(\Delta -2)!}{(\Delta+1)!} C^{(2)}_{\Delta-2}(\cos\theta_k)\,.
\end{equation}
This $SO(N)$ result is identical to the $SU(N)$ case obtained in~\cite{Giombi:2006de}. The D5-brane worldvolume has topology $AdS_2\times S^4$ with $AdS_2$ in the $AdS_5$ part of the background geometry and $S^4$ in the $\mathbf{RP}^5$ part. Since $\theta_k<\pi/2$, the $S^4$ we consider in this case is the same as the $S^4$ embedded in $S^5$ determined by $\theta=\theta_k$ in the parametrization given in~(\ref{fivesphere}). Thus the computations of the coupling of the supergravity modes to the $D5$-brane is the same as the $SU(N)$ case. 
Though the expressions of $T_{D5}$ in terms of $\lambda$ and $N$ for the $SO(N)$ case is different from the $SU(N)$ case.
In the $SO(N)$ case we have $T_{D5}=\frac{N}{8\pi^4}\sqrt{\frac{\lambda}{2}}$, while for the $SU(N)$ case the relation is $T_{D5}=\frac{N\sqrt{\lambda}}{8\pi^4}$. Taking this change and the relation  $c_{\rm SO}=\sqrt{2}c_{\rm SU}$ into account, we arrive at the conclusion that the OPE coefficients in the $SO(N)$ case is the same as the ones in the $SU(N)$ case. 
Thus in the $k\ll N$ limit, the relation $C_{A_k, \mathcal{O}}=k C_{\Box, \mathcal{O}}$ remains the same as in the $SU(N)$ case.
This can be obtained from the result $\theta_k^3\sim 3\pi k/2N$ in this limit and
\be C^{(2)}_{\Delta-2}(1)=\frac{(\Delta+1)!}{6(\Delta-2)!}\,.\ee

\section{OPE of Wilson loops in the spinor representation\label{spinor}}
Now we turn to the half-BPS circular Wilson loop in the spinor representation $S$ of $SO(N)$,
\be W_{S}[C]=\frac{1}{{\rm dim} S}  {\rm Tr}_{S} \mathcal{P} \exp \left[\oint_C i\Big(A_\mu(x) \dot{x}^\mu+{\rm i}\Theta_i \Phi^i(x)\Big){\rm d}s\right]. \ee
The dual description of such Wilson loop is in term of D5-brane whose worldvolume has topology $AdS_2\times \mathbf{RP}^4$ \cite{Witten:1998xy}. If we still chose the $\Phi^I$ to be  $\Theta^I=(1, 0, \cdots, 0)$. The embedding of the D5-brane is given by $u=0, \theta=\pi/2$ in the coordinates used in the previous section~\cite{Giombi:2020kvo}. In this case, the field strength of the worldvolume $U(1)$ gauge field vanishes. Taken into account the boundary terms, the total on-shell action of this D5-brane is
\be S^{\rm D5}=-\frac{N}{3\pi} \sqrt{\frac{\lambda}{2}}, \ee
so the holographic prediction for the expectation value of the Wilson loop in the spinor representation is~\cite{Giombi:2020kvo}
\be \langle W_S \rangle =\exp\left( \frac{N}{3\pi}\sqrt{\frac{\lambda}{2}}\right).\ee

As observed in~\cite{Giombi:2020kvo}, $F_{\psi\zeta}$, given by \eqref{Fpsizeta}, vanishes when $\theta_k=\pi/2$.
A shortcut to compute the OPE coefficient $C_{S, \mathcal{O}}$ using the result obtained in the previous section
is by setting $\theta_k=\pi/2$ in $C_{A_k, \mathcal{O}}$ and dividing the result by $2$ to take into account the change of D5-brane worldvolume from $AdS_2 \times S^4$ into $AdS_2\times \mathbf{RP}^4$,
\begin{align}
C_{S, \mathcal{O}}&=\frac{1}{2}C_{W_{A_k}, \mathcal{O}} \Big |_{\theta_k=\pi/2}
= \frac{2^{\Delta/2}}{3\pi}\sqrt{\Delta \lambda}
\frac{6(\Delta -2)!}{(\Delta+1)!} C^{(2)}_{\Delta-2}(0)  \nn\\
&=\frac{(-2)^{\Delta/2-1}\sqrt{\Delta\lambda}}{\pi(\Delta^2-1)}\,.
\end{align}
Here we have used that, for even $\Delta$,
\be C_\Delta^{(2)}(0)=(-1)^{\Delta/2}\frac{\Delta+2}{2}, \ee
obtained from the following generating function of the Gegenbauer polynomials $C_\Delta^{(\lambda)}(x)$, 
\be \frac{1}{(1-2xt+t^2)^\lambda}=\sum_{\Delta=0}^\infty C_{\Delta}^{(\lambda)}(x)t^\Delta. \ee

\section{Conclusion}

In this paper we studied the holographic duality of the ${\mathcal{N}}=4$~$SO(N)$ SYM theory and the Type IIB string theory on the
$AdS_5\times \mathbf{RP}^5$ background in the large $N$ and $\lambda$ limit.
To this end, we investigated the OPE coefficients of half-BPS circular Wilson loops in various representations.
The Wilson loop can be expanded in terms of local operators when the probing distances are much larger than the size of the Wilson loop. The coefficients can be extracted from the expansion for the operators we consider. Our focus is on the half-BPS CPOs and their corresponding gravity duals.
Specifically, we computed the correlation functions of local CPOs and the Wilson loop in the fundamental representation, the symmetric representation, the anti-symmetric representation and the spinor representation.
We studied the $SO(N)$ Wilson loops in the symmetric/anti-symmetric representations through their dual D3/D5-brane descriptions.
The appearance of Wilson loops in the spinor representation is a new feature in the $SO(N)$ theories.
In addition, we discussed the WH loops in the symmetric representation using a D3-brane with both electric and magnetic charges.
The $\mathcal{N}=4$ SYM theory with gauge group $SO(N)$ has some features different from the $SU(N)$ theory. We compared our results with the $\mathcal{N}=4$ $SU(N)$ SYM theory.

\section*{Acknowledgments}

This work is supported in part by the National Natural Science Foundation of China under Grants  No.~11975164,  No.~11935009, and Natural Science Foundation of Tianjin under Grants No.~20JCYBJC00910 and No.~20JCQNJC02030.

\appendix
\section{The coefficient $c$ of the bulk-to-boundary propagators}\label{appendix}
In this appendix, we compute the coefficient $c$ of the bulk-to-boundary propagator of the modes $s^I$. 
The action for the $s^I$, obtained from the full ``actual'' action of IIB supergravity~\cite{DallAgata:1997gnw} is~\cite{Lee:1998bxa} 
\begin{equation}
    S=\int_{AdS_5} {\rm d}^5 x \sqrt{{\rm det}(g_{AdS_5})} \frac{1}{2} B_I \left[ \partial_\mu s^I \partial^\mu s^I+\Delta (\Delta-4) (s^I)^2\right]\,,
\end{equation}  
where $B_I$ is given by
\be B_I=\frac{16}{\kappa^2} \frac{\Delta(\Delta-1)(\Delta+2)}{\Delta+1}z(\Delta)\,, \ee
where $\kappa$ is the coupling constant of type IIB supergravity, $z(\Delta)$ will be explained soon. 
Using 
\be 2\kappa^2=(2\pi)^7 g^2\alpha^{\prime 4}\,, \ee
and the relations $ \alpha^\prime= \sqrt{2/\lambda}$ and $ g_s= g_{\rm YM}^2 / (8\pi)$
for the $SO(N)$ case, we obtain
\be \kappa^2=\frac{(2\pi)^5}{8N^2}\,, \ee
which is same as the one in the $SU(N)$ case. 
$z(\Delta)$ is defined by 
\be \int_{\mathbf{RP}^5} {\rm d}^5y\sqrt{{\rm det}(g_{\mathbf{RP}^5})} Y^I Y^J =\delta^{IJ}z(\Delta)\,.\ee
The result of $z(\Delta)$ is
\be z(\Delta)=\frac{\pi^3}{2^\Delta (\Delta+1)(\Delta+2)}\,,\ee
which equals to half of the result in the $SU(N)$ case since the integration is over $\mathbf{RP}^5=S^5/\mathbf{Z}_2$.
Using the above result, we obtain
\be B_I=\frac{2^{2-\Delta}N^2\Delta (\Delta-1)}{\pi^2 (\Delta+1)^2}\,. \ee
The coefficient of the bulk-to-boundary propagator is 
\be c=\sqrt{\frac{\alpha_0}{B_I}}\,, \ee
where~\cite{Berenstein:1998ij} \be\alpha_0=\frac{\Delta-1}{2\pi^2}\,,\ee
which is identical for both $SO(N)$ and $SU(N)$ cases. 
Finally we arrive 
\be c=\frac{\Delta+1}{2^{(3-\Delta)/2} N\sqrt{\Delta}}\,, \ee
which equals $\sqrt{2}$ times the result in the $SU(N)$ case.

\providecommand{\href}[2]{#2}
\begingroup\raggedright\endgroup

\end{document}